\documentstyle[12pt,aaspp]{article}

\begin{document}

\title{Galaxy Aggregates in the Coma Cluster}

\author{Christopher J. Conselice$^1$ and John S. Gallagher III}

\affil{Department of Astronomy, University of Wisconsin, Madison, 475 N. Charter St., Madison, WI., 54706}

\altaffiltext{1}{chris@astro.wisc.edu}

\begin{abstract}

We  present evidence for a new morphologically defined form of small-scale 
substructure in the Coma Cluster, which we call galaxy aggregates.
These aggregates are dominated by a central galaxy, which is on
average three magnitudes brighter than the smaller aggregate members,
nearly all of which lie to one side of the central galaxy.  We have
found three such galaxy aggregates dominated by the S0 galaxies RB
55, RB 60, and the star-bursting SBb, NGC $4858$.

RB 55 and RB 60 are both equi-distant between the two dominate D galaxies 
NGC 4874 and NGC 4889, while NGC 4858 is located next to the larger E0 galaxy
NGC 4860.  All three central galaxies have redshifts consistent with Coma 
Cluster membership.   We describe the spatial structures of these unique 
objects and suggest several possible mechanisms to explain their origin.  These
include: chance superpositions from background galaxies,
interactions between other galaxies and with the cluster gravitational
 potential, and ram pressure.  We conclude that the most probable scenario of
creation is an interaction with the cluster through its potential.

\end{abstract}

\section{INTRODUCTION}
    Due to its  proximity and high density, the Coma Cluster is a good
location to investigate evolutionary effects in galaxy clusters, and for 
deriving
certain cosmological parameters (Crone et al. 1996; Dutta 1995).   Coma is
the closest rich cluster (Abell class 2) that can be studied in
detail, and considerable attention has been directed towards
understanding its nature, and deciphering its structure.
The model of Coma and other galaxy clusters as simple, virialized 
systems has been overthrown in the last few decades after the 
detection of significant substructure in a large number of clusters (e.g. West
1994). These features indicate that clusters have not had time 
to fully relax, and are likely to still be in the process of formation. 

The Coma cluster is now recognized as one
of the best examples of a cluster with  substructure, and several
method have been used to show that this substructure exists.

 The Coma cluster contains sub-condensations visible
in X-Rays first described in the work of Johnson et al. (1978).
Later studies showed that the X-Ray distribution is centered around
two point sources in the cluster (Davis \& Mushotzky, 1993; Sarazin,
1986).  As observing techniques and satellites  improved, more  X-Ray 
inhomogeneities were detected (Briel et al. 1992; White et al.
1993; Vikhlinin et al. 1997).  A recent study using redshifts found
that the velocity distribution of Coma members can be divided into
two components  centered around NGC 4874 and  NGC 4889 (Colless \&
Dunn, 1996).  This velocity difference is interpreted as the result
of the merger of two  clusters.   Evidence also exists for
sub-concentrations centered on the most massive galaxies (Meillier
et al. 1988; Baier 1984).  It is therefore no longer possible to
think of Coma as a single relaxed physical system.

Given its current complex state, it would not be surprising 
if interactions are affecting individual 
galaxies in the Coma Cluster.  These interactions would occur either through 
collisions  with other cluster members or with structures within the cluster.  
In the past, the impact of collisions within clusters was thought to 
be minimal due to the high relative velocities between individual 
galaxies that are expected in a virialized galaxy cluster. However, 
the observations indicate that galaxy-galaxy interactions are 
relatively frequent in some regions of the Coma cluster. In addition 
evidence for damaging collisions within Coma 
in the form of a giant stellar debris arc has
recently been discovered (Trentham et al. 1998). If a cluster is not 
virialized, then strong local interactions can occur between 
galaxies that are just falling into the cluster, as well as from 
more global tidal effects (Valluri 1993, Henriksen \& Byrd 1996). 

In this paper we present images of yet another curious form of 
galaxy within the Coma cluster. These consist of galaxies surrounded 
by numerous small companions. We call these objects "galaxy 
aggregates" so as not to prejudice their physical interpretation. 
However, we argue that at least some galaxy aggregates result from 
the disruption of disk galaxies within the Coma cluster. 

Section 2 presents our new imaging observations of the
Coma Cluster.  In section 3 we describe the galaxy aggregates, 
present photometry for these objects, and
quantify their positions within the Coma Cluster. The possible 
origins of galaxy aggregates are reviewed in Section 4, and we suggest 
that such objects are likely to be found only in those clusters which 
are actively accreting field galaxies.

\section{OBSERVATIONS}

Our images of the Coma Cluster which contain examples of galaxy
aggregates were taken with the WIYN\footnote{The WIYN Observatory is a
joint facility of the University of Wisconsin-Madison, Indiana
University, Yale University, and the National Optical Astronomy
Observatories.} 3.5m, f/6.5 telescope located at Kitt Peak National
Observatory between the nights of May 31 to June 2, 1997.  The thinned
2048$^2$ pixel S2kB charged coupled device (CCD) produced images with a scale
of 0.2 arcsec per pixel. The images cover a field-of-view of 6.8 x 6.8
arcmin. Broad-band B and R images were obtained for six different
regions of the cluster. Exposure time are 900s for the B-band and 600s
for the R-Band. The seeing ranged from 0.7 to 1 arcseconds full width
at half maximum.

These data are part of a larger study of the influence of 
cluster environments on low luminosity galaxies (Gallagher, Han \& 
Wyse 1997).  During a preliminary 
examination of the images, it became clear that 
they contained remarkable systems consisting of several luminous 
condensations around medium-luminosity disk systems.

Two of our galaxy aggregates are centered around the Coma S0
cluster members RB 55 and RB 60 (Rood \& Baum, 1967).   
The third galaxy aggregate is NGC 4858, 
whose central galaxy is a starbursting SBb member of 
the Coma cluster (Gallego et al. 1996).  

In addition to our morphological studies, 
aperture photometry was performed on the aggregates
using the IRAF package apphot. These were placed on the Landolt 
BR magnitude system using observations of his equatorial standard
star fields (Landolt 1992).

\section{DEFINITION AND MORPHOLOGY}

Galaxy aggregates have distinctive appearances in our images. They therefore 
at least merit a morphological definition, even if they turn out not to 
be a unique physical class of galaxy system. We define a ``galaxy aggregate" 
as follows:

\begin{enumerate}
 \renewcommand{\theenumi}{(\arabic{enumi})}

\item A galaxy aggregate consists of a primary lenticular 
(S0) or spiral galaxy that contains at least 5 distinct knots distributed asymmetrically 
within 2-3 optical radii of the primary. 

\item The primary of a galaxy aggregate is not a dominate  
member of a galaxy cluster, such as a cD or D galaxy. This is to 
avoid confusion with central galaxies in clusters which often are 
surrounded by dwarfs (e.g. in the Coma cluster).

\item The number of knots around the primary of an aggregate must 
be statistically in excess over that seen in the surrounding 
region.


 \end{enumerate}

\section{PROPERTIES OF COMA AGGREGATES}

The two galaxy aggregates associated with RB~55 (Figure 1 \& 2) and RB~60 (Figure 3) are located
between the two central D galaxies in Coma, NGC~4874 and
NGC~4889.  Both of these S0 primary 
galaxies appear to be embedded
in regions containing several compact, high surface brightness knots.
The average
separation of the knots from the central galaxies for both RB~55 and 
RB~60 are around  4h$^-$$^1$ kpc. 
The redshifts of RB~55 and  RB~60 are 7905 km~s$^-$$^1$ and
9833 km~s$^-$$^1$ respectively, 
where the mean redshift for the Coma cluster is 7200
km~s$^-$$^1$ (Colless \& Dunn, 1996).   The higher radial velocity
aggregate, RB~60, is more than 1$\sigma$ above the mean Coma velocity
dispersion, and could therefore be in-falling into the cluster core.   Ulmer et al. (1994) find dwarf galaxies to group in the Coma cluster, and suggest that RB 60 is a possible effect of this clumping.

A third object which morphologically qualifies as a galaxy
aggregate is the Coma Cluster member, NGC~4858 (Figure 4), a starburst SBb
galaxy located 13h$^-$$^1$ kpc away from the larger E0 galaxy NGC~4860.  
The redshifts for these members are 9436 km~s$^-$$^1$ and
7864 km~s$^-$$^1$ respectively.   Towards the interface between 
NGC~4858 and NGC~4860 are knots or dwarf galaxies distributed almost
symmetricaly about NGC~4858 (Figure 3).  The average distances of
the knots from NGC~4858 is 6h$^-$$^1$ kpc.

The luminosity function of the aggregates are shown in Figure 5.
RB~55 and RB~60 are both dominated by a galaxy of R magnitude
around 17 with NGC 4858 aggregate having a central galaxy of
magnitude 18.6.  The NGC~4858 and RB~55 aggregates have similar luminosity
functions, which rise at fainter magnitudes.  RB~60 has a
luminosity function which peaks between 22 and 22.5 and falls off
at fainter magnitudes.

The (B-R) vs. R plots (Figure 6) show that two of the
aggregates contain knots covering a wide range in brightness, while those 
in NGC~4858 are near our faint limits. The colours of the knots in NGC~4858 
and RB~60 are very red, while those in RB~55 are bluer and 
close to the colour of the primary galaxy and have similar 
colours to the dE galaxies in the Coma cluster (Secker 1996).

Images of the aggregates RB 55 and RB 60 do not show significant amounts of H{$\alpha$} in the smaller members.

\section{RESULTS AND ANALYSIS}

\subsection{Chance Alignments}

In a  dense cluster like Coma, there is a  possibility that 
galaxy aggregates result from chance alignments of galaxies at different
distances. To test this possibility, the number density of galaxies
detected in several fields in Coma, and the densities of the two
aggregates are computed.  The mean number density  of detected galaxies in
Coma cluster away from the aggregates on our 6.7~arcmin field 
CCD images is 4.7
galaxies arcmin$^-$ $^2$.  The surface density of 
knots in the aggregates is 36 galaxies arcmin$^-$ $^2$ for RB~55, 32 galaxies per arcmin$^-$ $^2$ for RB~60 and 38 galaxies per arcmin$^-$ $^2$ in NGC 4858.  
The galaxy aggregates are effectively 6~$\sigma$ fluctuations, which 
have less than a 0.01\% chance of occurring due to a random 
superposition of Coma galaxies. Further evidence against 
the random hypothesis is our failure to observe similar superpositions in other
regions of Coma, or in the clusters: Abell 2199, AWM 5,
AWM 3, or Perseus.   These aggregates are therefore
relatively unusual, whatever their origin may be.

Secker et al. (1997) consider galaxies with colours of (B-R) $>$ 2
or (B-R) $<$ 0.9 as either too red, or too blue to be dwarf 
members of the cluster, and rejected such objects as being 
in the background or foreground.  Based on these criteria, most of the 
knots in the NGC~4858 and RB~60 aggregates could be background 
galaxies.  Perhaps these are examples of foreground S0 systems 
superimposed on a distant poor galaxy clusters or groups?  We 
suspect that this is not a universal explanation as we do not 
find examples of such background clusters without a foreground 
object in our images.  Therefore, some special effect would have 
to be invoked to amplify background objects near the S0s. The 
one candidate for such a process, gravitational lensing, is unlikely 
on geometrical grounds; the images are neither distorted along 
arcs nor at small angles from the nuclei of the S0 primaries. 

\subsection{Gravitational Interactions}

 The morphologies of the Coma galaxy aggregates are suggestive of a 
central galaxy surrounded by dense knots of stars, which could have 
been produced by recent unusual events, such as a first passage 
through the Coma cluster.  Under these conditions star formation 
in the disk of the central galaxy or in surrounding dwarf satellites 
might be triggered via interactions with other cluster members, or 
as an effect of the clusters overall gravitational potential. 
Evidence for star formation induced by interactions within clusters 
comes primarily from the presence of blue disk galaxies responsible 
for the Butcher-Oemler effect seen most prominently in moderate 
redshift galaxy clusters (e.g., Dressler \& Gunn 1983, 
Lavery \& Henry 1988, 1994, Couch et al. 1994). A high fraction of 
these blue galaxies show signs of recent gravitational interactions with 
nearby galaxies, and have the expected kinematics of recent infalls 
into their clusters. 

Despite its proximity and the dominance of early-type galaxies in its core, 
the Coma cluster also contains a population of blue 
galaxies (Bothun \& Dressler 1986, Caldwell et 
al. 1993 and references therein). Thus, this cluster may also contain 
galaxies that are responding to interactions within the cluster. The 
only likely binary aggregate pair is NGC~4858 whose nearby E companion 
NGC~4860 is at a 1572~km~s$^{-1}$ lower radial velocity. Such large 
velocity differences will reduce the severity of galaxy-galaxy 
collisions, but they may still be able to produce a significant 
starburst as is seen in NGC~4858 (Moore et al. 1996). 


A more general mechanism for perturbing infalling galaxies and 
producing unusual stellar clumps may come from interactions with 
the clusters overall tidal field.
In Merritt's (1984) model, galaxies which are close to the cluster core
will experience maximal tidal forces from the cluster potential.  
The tidal disruption experienced by a galaxy depends on the velocity
dispersion of the galaxy $v_g$ and that of
the cluster $v_{cl}$. The tidal radius $r_T$ then varies with the cluster 
core radius $R_c$,  
$$r_T=R_c\frac{1}{2}\frac{v_ g}{v_{\rm cl}}. $$
Galaxies whose disk or satellite system radii exceed $r_T$ may experience 
significant distortion from the cluster tides.

In Coma $R_c=0.15^\circ$ and $v_{\rm cl}=1062\,{\rm
km\,s^{-2}}$ (Kent \& Gunn 1982).  
Adopting 100 km~s$^{-1}$ for the equivalent velocity dispersion of a 
moderate luminosity S0 galaxy, the
condition for tidal disruption at the Coma cluster core 
becomes $r_T>$8~kpc. In the RB~55 and RB~60 systems the faint knots 
are found at about this radius.

A possibly more attractive way the cluster could produce an
aggregate is as a result of the reaction of galaxy disk 
to external tidal forces. This process may lead to 
an epoch of enhanced star formation activity as well as thickening 
of the stellar disk, as discussed by Valluri (1993) and 
Henriksen and Byrd (1996). However, none of these models lead in a 
transparent way to the production of knots of stars which are the 
morphologically defining characteristic of galaxy aggregates. 

\subsection{Ram-Pressure Stripping}

Ram-pressure is the second primary type of environmental influence 
that a rich cluster may exert on its members.  This effect was initially 
described by Gunn \& Gott (1972) as a means to convert spirals into 
S0 galaxies within rich galaxy clusters. While there is now extensive 
observational evidence that gas is stripped from spirals in galaxy 
clusters (e.g., Vigroux et al. 1986, Haynes \& Giovanelli 1986) 
the details of this process are still debated (see Nulsen 1982, 
Gaetz, Salpeter, \& Shaviv 1987, Henriksen \& Byrd 1996). 

Certainly the copious hot intracluster medium in the Coma cluster 
will have an effect on its member galaxies (White et al. 1993). 
However, whether this also can yield clump formation, as in the 
case of tidal interactions, also remains uncertain. 

\section{DISCUSSION}
Three galaxy aggregates centered around NGC 4858, RB 55 and RB 60
have been found in the Coma cluster.  
These consist of moderate-sized  disk galaxies 
nested in several luminous knots or dwarf galaxy-like objects,
which in RB~55 and RB~60 
are {\it not} emission line regions.
The number densities of objects in the aggregates 
are seven times larger than for small objects 
in surrounding regions of the Coma cluster.  The 
large over-densities in the aggregates  
and the absence of galaxy aggregates in other nearby galaxy 
clusters suggest that they are statistically unlikely to
be chance alignments. However, projections of background objects
possibly combined with weak gravitational lensing cannot be excluded.

The most likely
model for the creation of the knots or dwarf galaxies 
in aggregates is through gravitational interactions with 
neighboring galaxies or the cluster as a whole during the infall 
of disk galaxies into the cluster. Tests of possible models 
require spectroscopy to determine the stellar content, kinematics, 
and redshifts of the knots within the aggregates. If the aggregates 
are background objects, then this will be immediately clear from 
their redshifts, while if, as we believe, they are produced by 
disturbances in a disk galaxy or its companions, then their radial 
velocities will be close to those of the primary galaxy, and the 
spread in velocities will be $\approx$300~km~s$^{-1}$ or less, as 
is typical of internal motions in small galaxies. In this model 
the presence of galaxy aggregates in the Coma cluster would then 
be another indication of the dynamic evolutionary state of this 
cluster.

\section*{ACKNOWLEDGMENTS}

We thank D.J. Pisano and Keivan Stassun for assistance with the
photometry.  This work has made use  of the NASA/IPAC Extragalactic
Database(NED).

\newpage

\vspace{3cm}

FIGURE CAPTIONS:

Fig.1 --- The Coma field with the Aggregate RB 55 at centre.  The high surface brightness, almost linear knots, close to the central S0 stand out as being unique. The size of the image is 4.3 arcmin on each side.

Fig.2 --- A close up view of Aggregate RB 55. The image is 1' on each side, with a scale of 0.2'' per pixel.

Fig.3 --- The aggregate RB 60.  This 1' by 1' image shows the large central S0 galaxies surrounded by smaller dwarf galaxies or semi-stellar knots.

Fig.4 --- The aggregate NGC 4858.  This aggregate is unique from the other two in that it is close to another galaxy, NGC~4860.  The knots in this image are all redder than NGC 4858.  The image size is 1' by 1'.

Fig.5 --- The luminosity function for the Aggregates.
The magnitudes for NGC 4858, RB 55 and RB 60 are 18.6, 17.0, and
17.4.   The luminosity function for NGC 4858 and RB 55 both rise at
fainter magnitudes, whereas the RB 60 function peaks near magnitude
22.

Fig.6 --- The colour (B-R) vs. R relation.  The colours
of the central galaxies are (B-R)=0.9, 1.4, and 1.5, these are
denoted on the figure as a dashed line.  For NGC 4858 and RB 55 a
slight trend towards bluer colours at fainter magnitudes is
detected.  The colours of the objects in RB 60 tend to be a bit
redder than the central galaxy but have no particular trend.  The
colours and magnitudes for NGC 4858 are similar and the RB 55
colours are almost all bluer than the central galaxy.

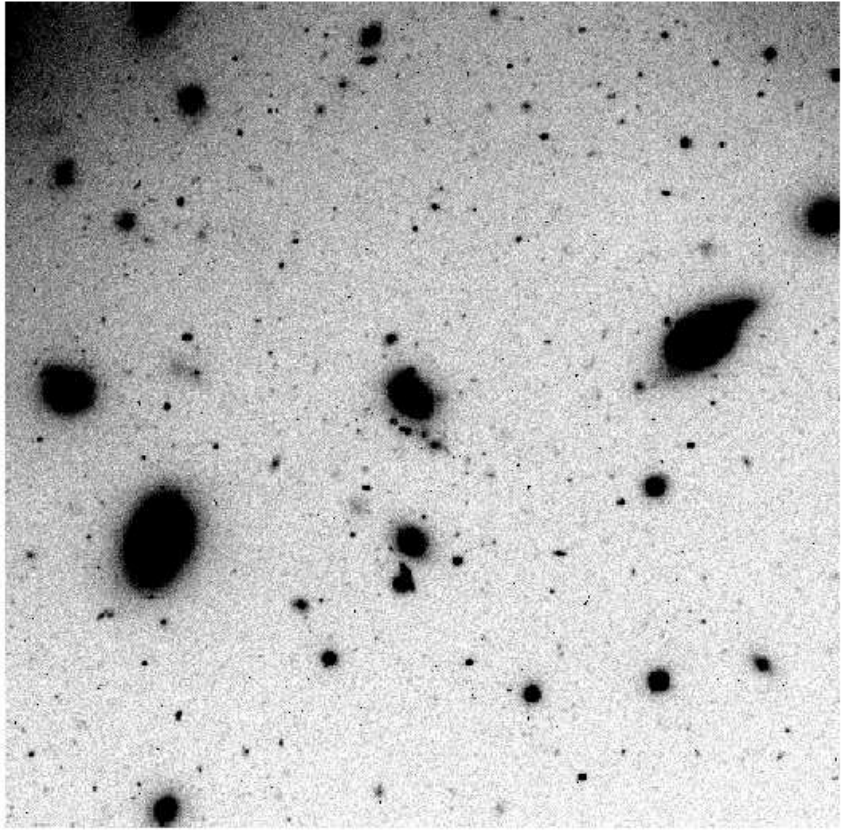
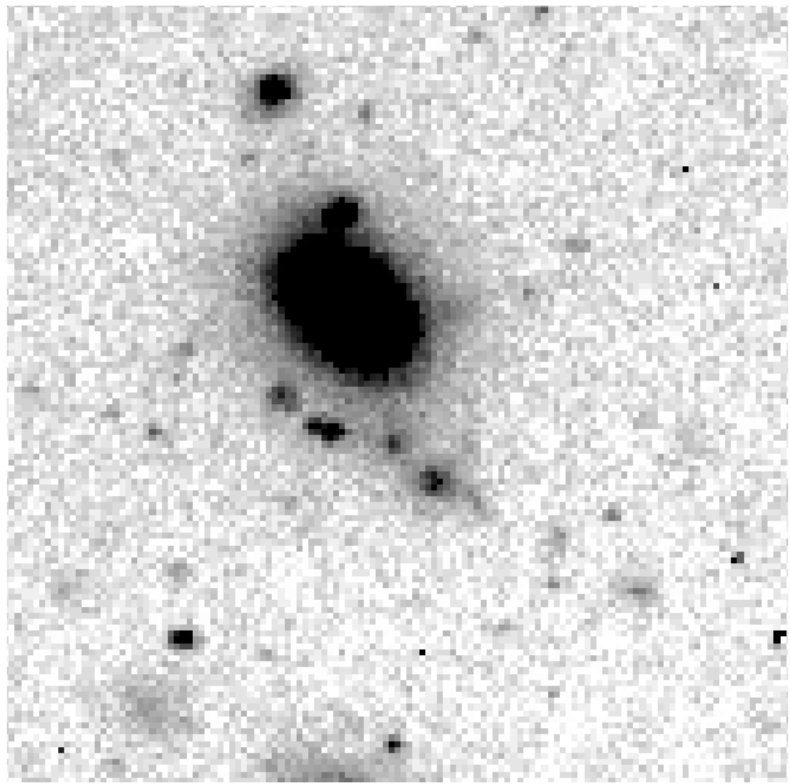
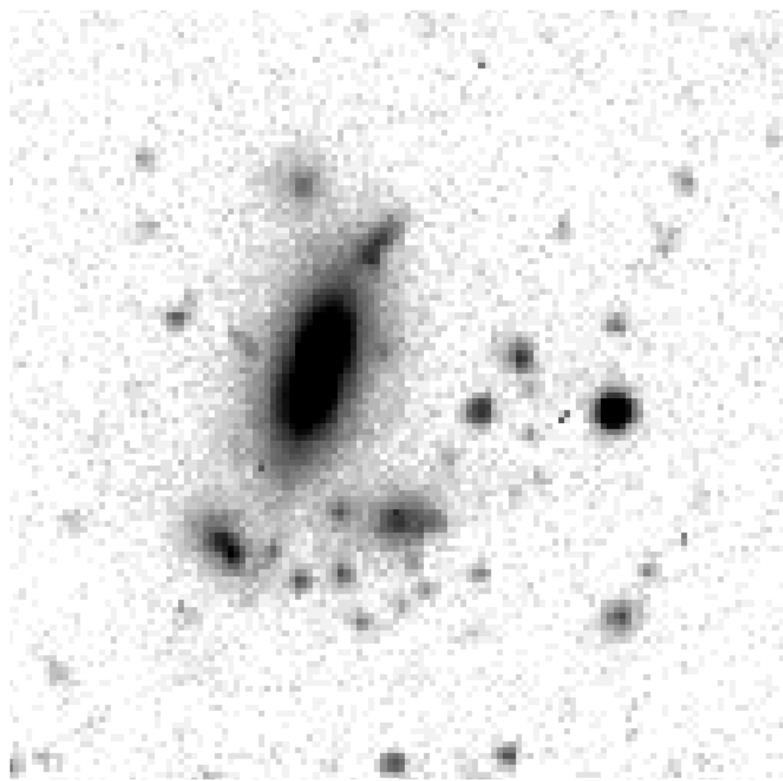
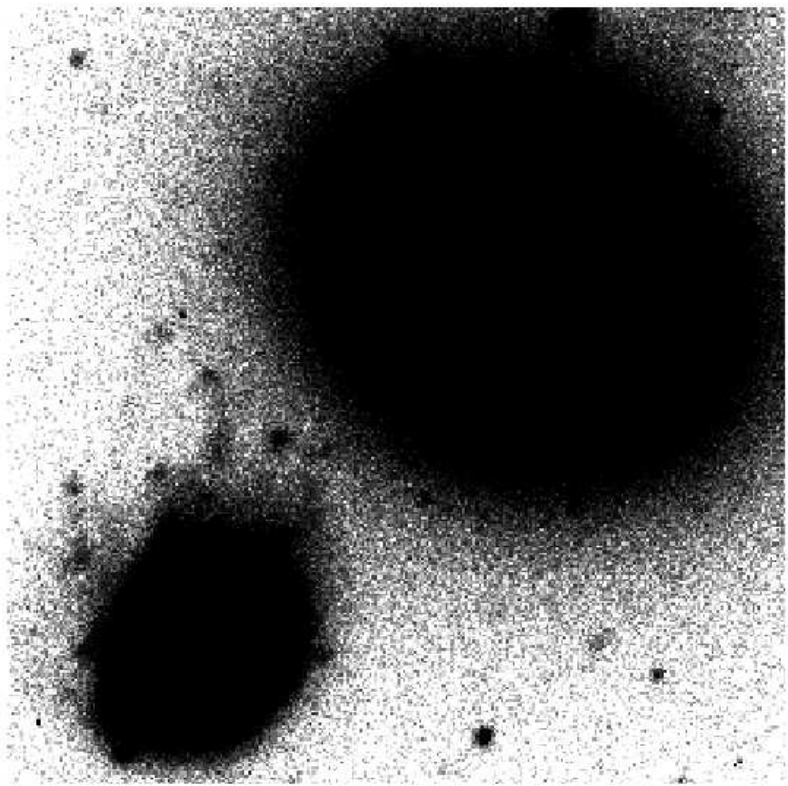
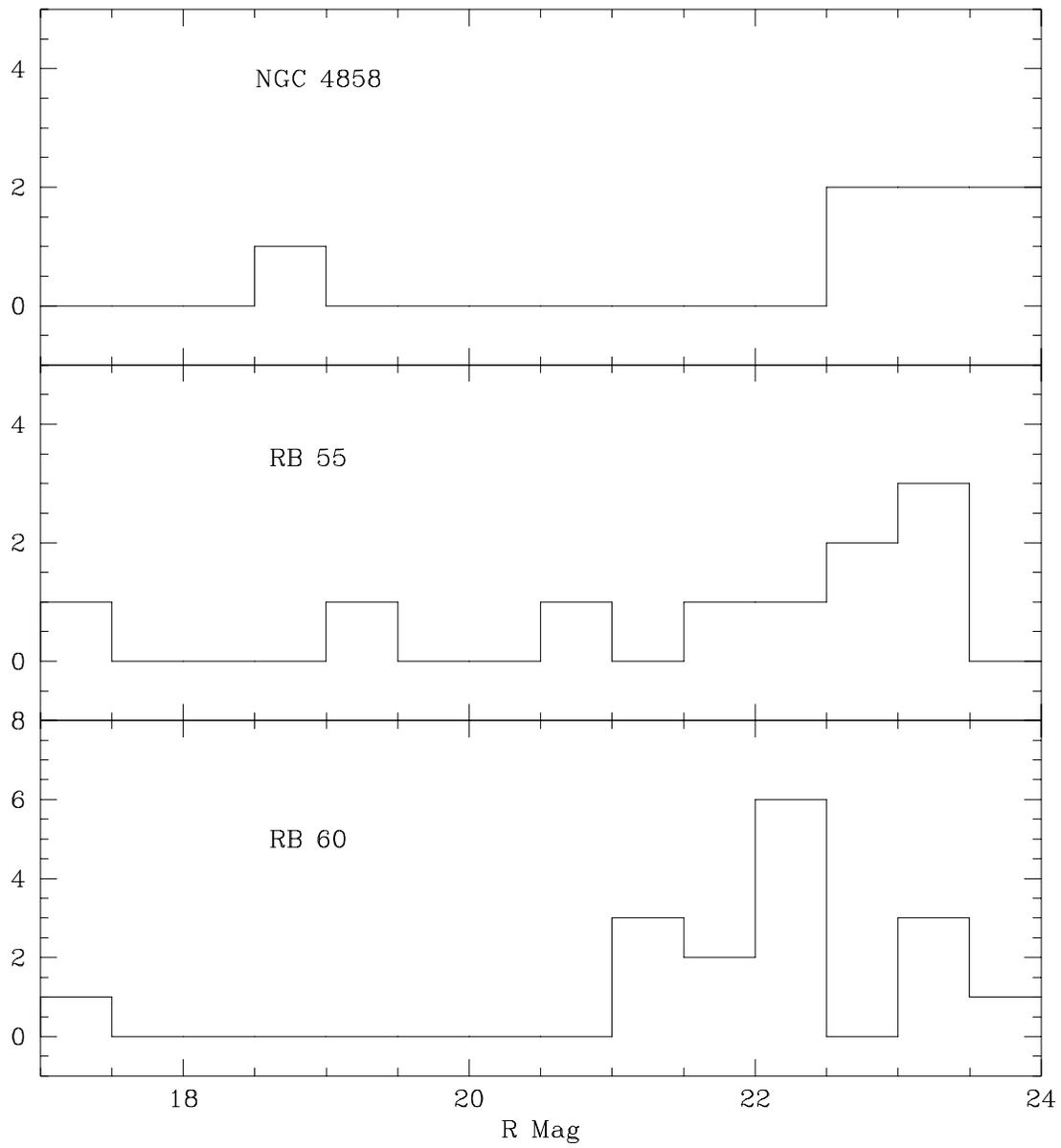
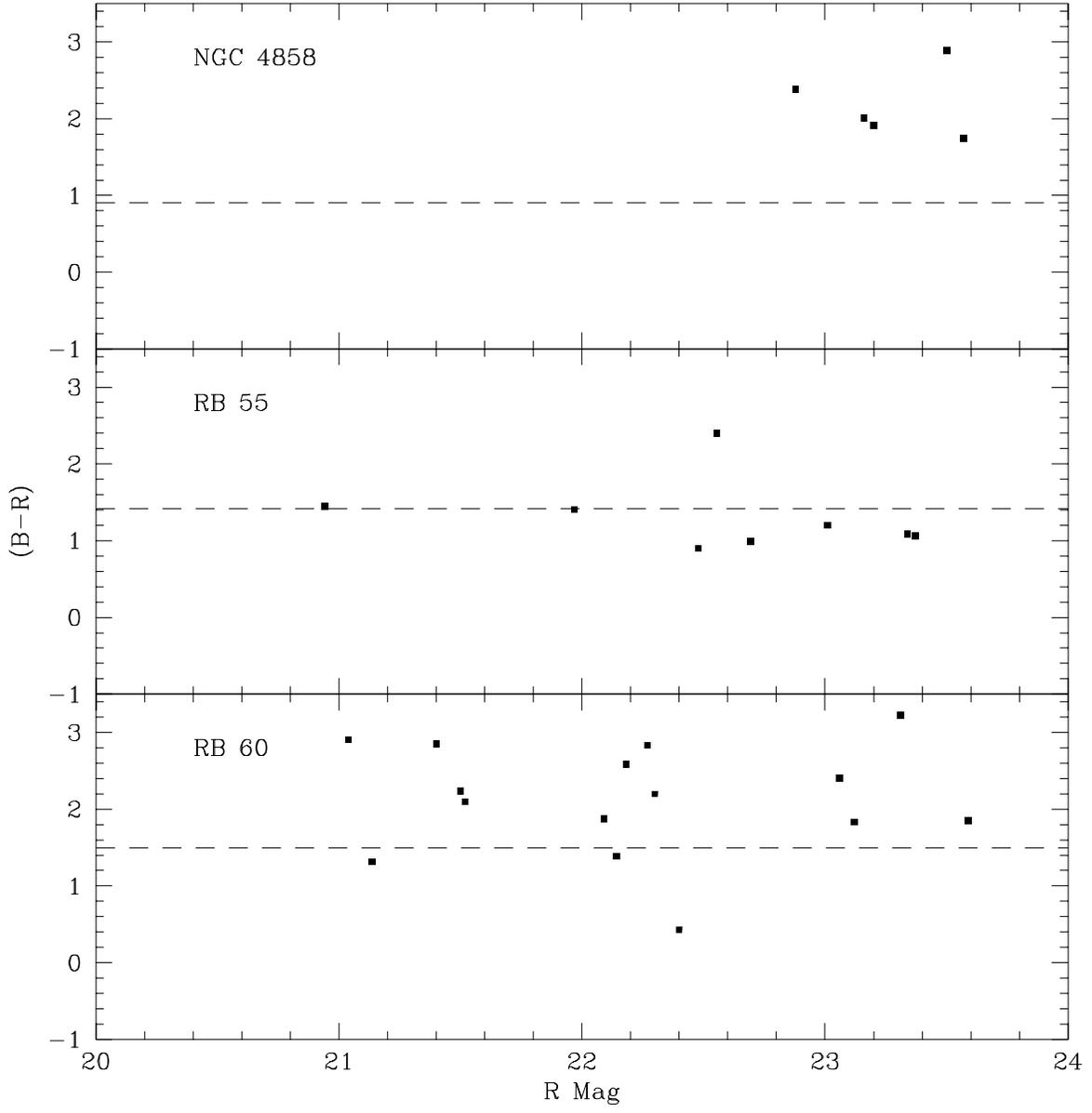
\end{document}